\documentclass[lettersize,journal]{IEEEtran}
\usepackage{amsmath,amsfonts}
\usepackage{algorithm}
\usepackage{array}
\usepackage[caption = false, labelformat=simple]{subfig}
\usepackage{textcomp}
\usepackage{stfloats}
\usepackage{url}
\usepackage{verbatim}
\usepackage{graphicx}
\usepackage{cite}
\usepackage[numbers]{natbib}
\usepackage[colorlinks=true, linkcolor=black, citecolor=black, urlcolor=black]{hyperref}
\usepackage{booktabs}
\usepackage{multirow}
\usepackage{cleveref}
\usepackage{xspace}
\usepackage{tablefootnote}
\usepackage{algpseudocode}
\usepackage{listings} 
\usepackage{xcolor} 
\usepackage{fancyvrb} 
\usepackage{lettrine}
\usepackage{orcidlink}

\lstdefinestyle{mypython}{
    language=Python,
    basicstyle=\ttfamily\footnotesize, 
    keywordstyle=\color{blue},
    commentstyle=\color{gray},
    stringstyle=\color{red},
    showstringspaces=false,
    frame=single,
    breaklines=true,
    postbreak=\mbox{\textcolor{red}{$\hookrightarrow$}\space}
}

\lstnewenvironment{PythonB}[1][]{\lstset{style=mypython, frame=none, breaklines=true, #1}}{}

\crefname{table}{Table}{Tables}
\crefname{figure}{Figure}{Figures}
\crefname{section}{Section}{Sections}

\newcommand{\xr}[1]{\textcolor{black}{#1}}

\newcommand{\model}{\emph{AEFS}\xspace}
\newcommand{\modelFullName}{Adaptive Early Feature Selection\xspace}
\newcommand{\EmbRatio}{$\Delta \mbox{PaE}$\xspace}

\newcommand{\auxModel}{auxiliary model\xspace}
\newcommand{\mainModel}{main model\xspace}

\newcommand{\AuxModel}{Auxiliary Model\xspace}
\newcommand{\MainModel}{Main Model\xspace}

\newcommand{\EAL}{Embedding Alignment Loss\xspace}
\newcommand{\PAL}{Prediction Alignment Loss\xspace}

\def\ie{\textit{i.e.}~}

\def\etc{\textit{etc.}}
\def\etal{\textit{et al.}\xspace}

\begin{document}

\title{AEFS: Adaptive Early Feature Selection for Deep Recommender Systems}

\author{Fan Hu\IEEEauthorrefmark{1} \orcidlink{0000-0002-5371-7780}, Gaofeng Lu\IEEEauthorrefmark{1} \orcidlink{0000-0001-9619-5049}, Jun Chen, Chaonan Guo, Yuekui Yang, Xirong Li\IEEEauthorrefmark{2} \orcidlink{0000-0002-0220-8310},~\IEEEmembership{Member,~IEEE,}

\thanks{Received 20 November 2024; revised 15 May 2025, 11 August 2025; accepted 2 September 2025. 
Date of publication XX XXX 2025; date of current version XX XXX 2025. This work was supported in part by the National Natural Science Foundation of China under Grant 62576348 and Grant 62172420, and in part by the Tencent Marketing Solution Rhino-Bird Focused Research Program. Recommended for acceptance by XXX. \textit{(Corresponding author: Xirong Li.)}}
\thanks{Fan Hu and Xirong Li are with Renmin University of China, China (e-mail: hufan\_hf@163.com, xirong@ruc.edu.cn).}
\thanks{Gaofeng Lu is with  University of Science and Technology of China, China (e-mail: lugaofeng@ustc.edu). }
\thanks{Jun Chen, Chaonan Guo, Yuekui Yang are with  Tencent, China (e-mail: arthurjchen@tencent.com, thinkerguo@qq.com, yuekuiyang@gmail.com). }
\thanks{ $*$Equal contribution. 
}
\thanks{Digital Object Identifer xxx/TKDE.xxx}

}

\markboth{IEEE Transactions on Knowledge and Data Engineering,~Vol.~XXX, No.~XXX, Month~2025}%
{Hu \MakeLowercase{\textit{et al.}}: AEFS: Adaptive Early Feature Selection for Deep Recommender Systems}


\maketitle

\begin{abstract}

The quality of features plays an important role in the performance of recommender systems. Recognizing this, feature selection has emerged as a crucial technique in refining recommender systems. Recent advancements leveraging Automated Machine Learning (AutoML) has drawn significant attention, particularly in two main categories: \emph{early} feature selection and \emph{late} feature selection, differentiated by whether the selection occurs before or after the embedding layer.
The early feature selection selects a fixed subset of features and retrains the model, while the late feature selection, known as adaptive feature selection, dynamically adjusts feature choices for each data instance, recognizing the variability in feature significance.
Although adaptive feature selection has shown remarkable improvements in performance, its main drawback lies in its post-embedding layer feature selection. This process often becomes cumbersome and inefficient in large-scale recommender systems with billions of ID-type features, leading to a highly sparse and parameter-heavy embedding layer.
To overcome this, we introduce \modelFullName (\model), a very simple method that not only adaptively selects informative features for each instance, but also significantly reduces the activated parameters of the embedding layer. \model employs a dual-model architecture, encompassing an \auxModel dedicated to feature selection and a \mainModel responsible for prediction. To ensure effective alignment between these two models, we incorporate two collaborative training loss constraints. Our extensive experiments on three benchmark datasets validate the efficiency and effectiveness of our approach. Notably, \model matches the performance of current state-of-the-art Adaptive Late Feature Selection methods while achieving a significant reduction of 37. 5\% in the activated parameters of the embedding layer. We believe that this work opens up new possibilities for feature selection.

\end{abstract}



\section{Introduction}

\begin{figure}
\centering
    \subfloat[Early feature selection (AutoField \cite{wang2022autofield}, AutoFSS \cite{wei2023automatic}, \etc) \label{fig:early_fs}]{\includegraphics[width=\columnwidth]{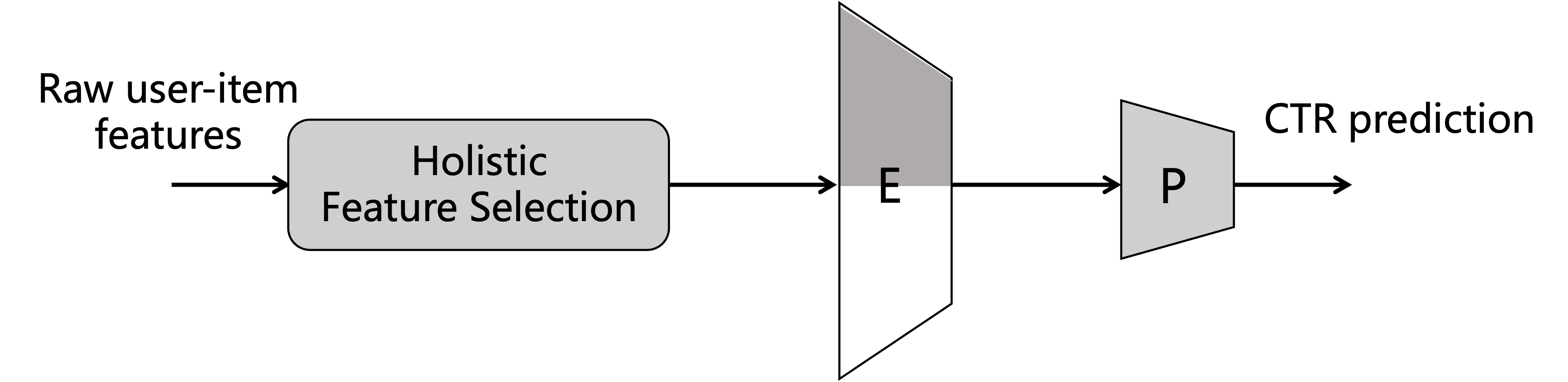}}

    \subfloat[Late feature selection (AdaFS \cite{lin2022adafs}, MvFS \cite{lee2023mvfs}, \etc)\label{fig:late_fs}]{\includegraphics[width=0.77\columnwidth]{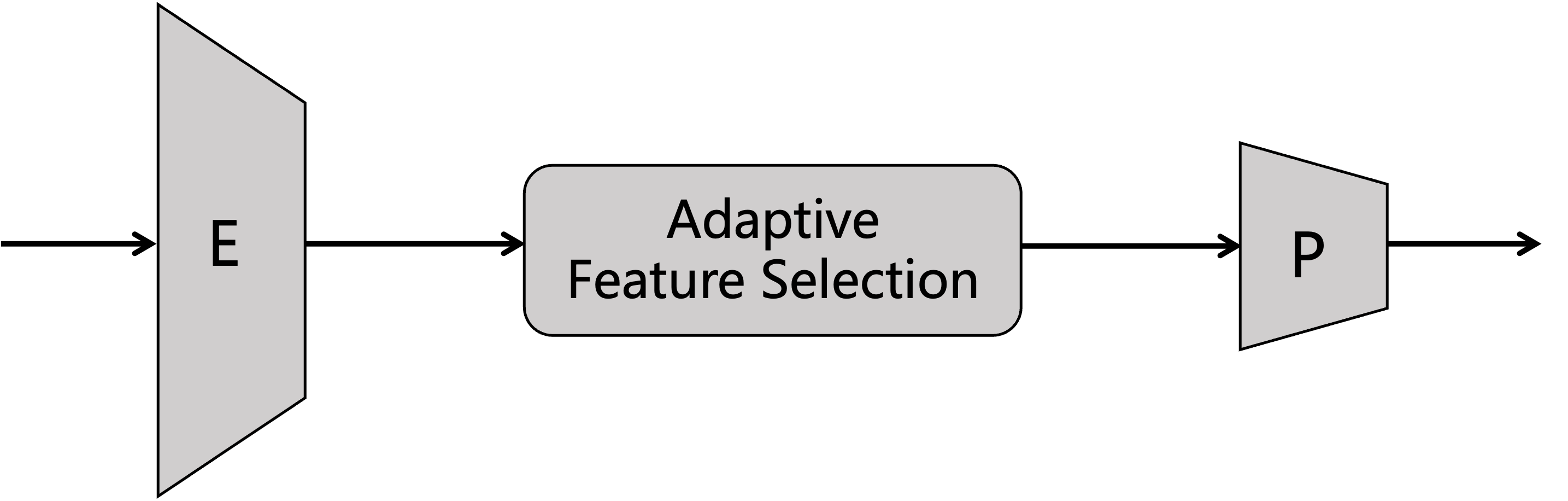}}

    \subfloat[Adaptive early feature selection (\emph{this paper}) \label{fig:near_early_fs}]{\includegraphics[width=0.77\columnwidth]{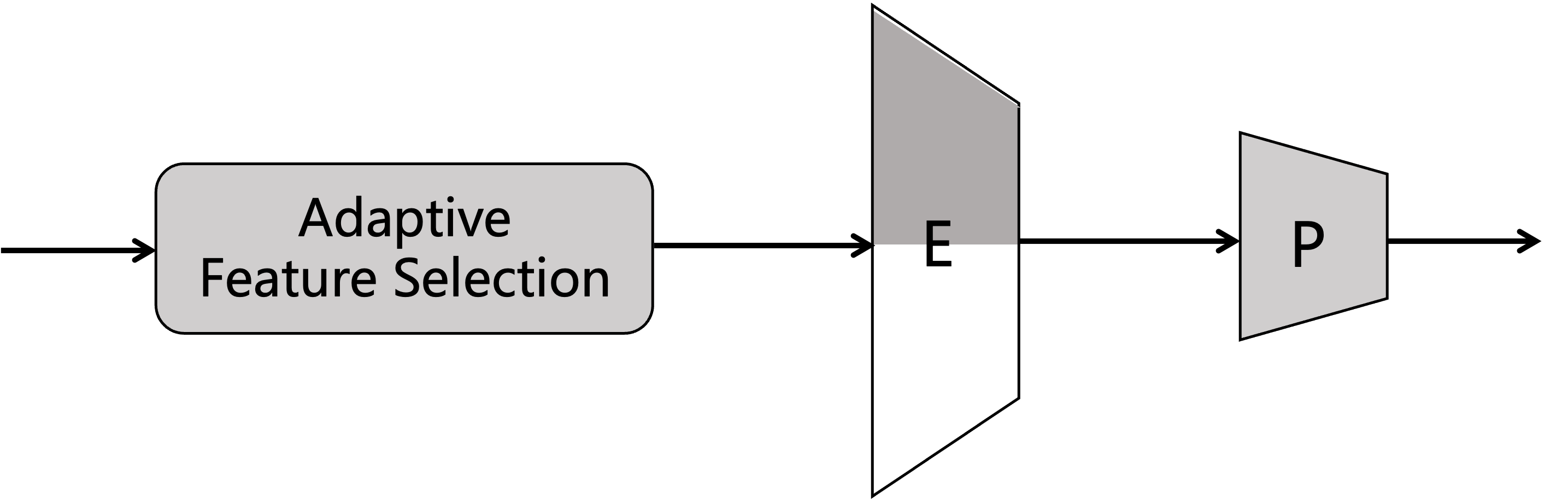}} 
    \caption{\textbf{High-level illustration of feature selection (FS) in a deep recommendation system (DRS)}.  Given raw user-item features as input, a DRS employs an embedding layer (E) to transform the input into an array of dense features vectors, which are then fed into a prediction layer (P) for click-through rate (CTR) prediction. 
    (a) Early FS, which occurs before the E layer,  effectively reduces the activated embedding parameters for forward and backward computation (the shadow reigon in E), yet at the cost of relatively lower prediction accuracy.
    (b) Late FS, being adaptive to the input, is more accurate, yet occurs after the E layer and thus practically uses its full parameters. (c) The proposed adaptive early FS (AEFS) combines the strengths of both: it dynamically selects different fields and substantially reduces the active embedding parameters.
    }
    \label{Figure:three_fs}
\end{figure}

\lettrine[lines=3]{D}{eep}  Recommender Systems (DRS) utilize deep learning to enhance prediction accuracy and personalization in recommendations. They differ from traditional systems by using complex neural networks, enabling them to capture intricate user-item relationships and process more abstract data representations. This capability allows DRS to offer more accurate and tailored recommendations to users \cite{wang2017dcn, guo2017deepfm, DERRD, HetComp, concf, TD}.
Practical large-scale DRS typically involves numerous categorical feature fields, encompassing user data (such as user ID, age and city), item specifics (such as category and item ID) and contextual information (time, location, user's purchase history, \textit{etc.}) \cite{zhu2020fuxictr, zheng2023automl}.  
For clarity, we use “feature field” to represent a class of features and “feature value” to represent a certain value in a specific feature field.
The features are first mapped into real-valued embeddings using an embedding layer. Subsequently, embeddings from the embedding layer are transformed to generate predictions through a prediction layer, which typically comprises a few fully-connected layers in practical implementations. Currently, the scale of recommender systems has reached the level of trillions of parameters \cite{mudigere2022software, lian2022persia}, with the embedding layer accounting for 99.99\% of these parameters. Extracting feature embeddings from this huge and sparse embedding layer, as well as updating the corresponding embedding layer parameters during backward propagation, is time-consuming in end-to-end training and inference \cite{lian2022persia}.

Not all features are equally important for user-item interactions \cite{lin2022adafs, wang2022autofield, lyu2023optimizing, lyu2022optembed}. Some features may exhibit redundancy or lack relevance, negatively impacting the model's accuracy and efficiency.
To address these challenges, recent advancements in automatic feature selection techniques \cite{wang2022autofield, wei2023automatic, lyu2023optimizing, lin2022adafs, lee2023mvfs, jia2024erase} have leveraged the power of Automated Machine Learning (AutoML) \cite{liu2018darts, luo2018NAS}. 
Jia  \etal propose a benchmark for feature selection and systematically compares eleven selection methods across multiple datasets \cite{jia2024erase}.
These techniques automatically identify and select the optimal features during the model training process, which shows effectiveness for DRS models.
As illustrated in Fig. \ref{Figure:three_fs}, we classify existing works into two distinct groups, depending on  the positioning of the feature selection process in relation to the embedding layer: \emph{early} feature selection, which involves selecting features before embedding, and \emph{late} feature selection, performing selection or re-weighting of features after embedding.


An early feature selection approach typically works in a two-stage search-\emph{and}-retrain style \cite{wang2022autofield, wei2023automatic, lyu2023optimizing}, see Fig. \ref{fig:early_fs}. In the first stage, all feature fields are input into a deep model and an importance scoring network or Network Architecture Search (NAS) methods are used to select $K$ optimal feature fields. Subsequently, only the selected feature fields are utilized to retrain the deep recommendation model. This approach aims to eliminate unimportant feature fields, thereby reducing the model activated parameters and improving performance. However, completely removing feature fields carries significant risks to the performance. Due to the long-tail distribution of features, some feature fields may overall appear less important but are crucial for certain user-item interactions. Subsequent studies replicating the AutoField \cite{wang2022autofield} experiment also indicate that when up to 50\% of the features are removed, AutoField, compared to the No Selection Baseline, does not show a significant improvement or slightly decreases in the AUC metric \cite{lin2022adafs, lee2023mvfs, lyu2023optimizing}.

The second category, \emph{late} feature selection (Fig. \ref{fig:late_fs}),
usually encompasses adaptive feature selection according to the information of feature embedding.  As initially introduced by Lin \etal. \cite{lin2022adafs}, AdaFS implements an adaptive selection strategy.  Unlike previous methods that focused on selecting a static, globally-fixed subset of feature fields, AdaFS dynamically selects feature embeddings tailored to each specific user-item interaction, thereby achieving notable enhancements in performance. Subsequent work such as MvFS \cite{lee2023mvfs} employs a multi-view network architecture composed of multiple sub-networks, enabling more effective selection of informative features for individual instances. However, a critical limitation of \textit{late} feature selection methods is that they apply feature selection after the embedding layer, which means that they do not substantially improve parameter efficiency.  { {Input-dependent feature selection before the embedding layer is thus crucial for an accurate and efficient DRS.}}

\begin{table}[t]
\centering \setlength{\tabcolsep}{3.0pt}
\caption{\textbf{Decomposition of trainable parameters in a DRS}, where the embedding layer (E) accounts for 99.99\% of the parameters. $N_{id}$ indicates the number of distinct feature IDs.  Embedding size: 32.}
\label{tab:parm_emb_dense}

\begin{tabular}{@{}l|l|l|l|l@{}}
\toprule
\textbf{Dataset} & \textbf{$N_{id}$} & \textbf{\#Params of E} & \textbf{Choice of P} & \textbf{\#Params of P} \\ \midrule
\multirow{3}{*}{Avazu} & \multirow{3}{*}{2,018,012} & \multirow{3}{*}{64.58M} & MLP \cite{zhang2016deep} & 0.0115M \\ 
 &  &  & DeepFM \cite{guo2017deepfm} & 0.0114M \\
 &  &  & DCN \cite{wang2017dcn} & 0.0164M \\ \midrule
\multirow{3}{*}{Criteo} & \multirow{3}{*}{1,086,810} & \multirow{3}{*}{34.78M} & MLP & 0.0202M \\
 &  &  & DeepFM & 0.0201M \\
 &  &  & DCN & 0.0289M \\
 \bottomrule
\end{tabular}

\end{table}

The primary aim of this paper is to develop an \modelFullName (\model) method.  As presented in Fig. \ref{fig:near_early_fs}, the proposed method aims to not only select relevant features before the embedding layer but also adaptively choose the optimal subset of feature fields for each user-item interaction, thereby enhancing the accuracy and efficiency of the DRS model. We introduce an \auxModel, which is designed to identify and generate feature importance scores for each user-item interaction. These scores are then utilized by the \mainModel to facilitate adaptive feature selection before the embedding layer.
For the \auxModel, we carefully chose a significantly smaller embedding dimension compared to those employed in the \mainModel. To ensure the coherence between the \auxModel and the \mainModel, two collaborative training loss constraints are employed, aligning the embeddings and prediction scores of both models. This study addresses the existing gap in the literature and presents an advancement in the context of Deep Recommender Systems. The proposed approach is expected to not only enhance the performance of DRS models but also reduce the activated parameter in the embedding layer, thereby optimizing both efficiency and effectiveness.


The contributions of this work are as follows:
\begin{itemize}
    \item  \xr{We propose a new categorization of existing FS methods, namely early FS and late FS. Such a categorization is essential for us to identify key limitations in the previous works. That is, early FS is suboptimal in terms of prediction accuracy, while the parameter-efficiency of late FS is questionable.}  
    \item We propose the Adaptive Early Feature Selection (\model) method, which combines the best aspects of the previous two methods.  By employing an \auxModel and two collaborative training loss constraints, \model optimally selects relevant features for each user-item interaction and significantly reduces embedding complexity.
    \item The \xr{viability} of the proposed method is \xr{verified} through comprehensive experiments \xr{on three public datasets}, showcasing improved efficiency and accuracy in recommender systems. AEFS is open-source at \url{https://github.com/fly-dragon211/AEFS}.
\end{itemize}

\section{Related work}\label{sec:related}

Feature selection plays a vital role in enhancing prediction accuracy and improving parameter-efficiency of a DRS. In what follows, we discuss feature selection in general, followed by feature selection techniques that are tailored to a DRS. Accordingly, we explain the novelty of the proposed \model.

\subsection{Feature Selection in General}
Feature selection has been a classic research direction since the era of traditional machine learning, and there are three main categories: the wrapper methods, the filter methods and the embedded methods \cite{guyon2003introduction, chen2022comprehensive}.
Wrapper methods evaluate different subsets of features to determine their impact on the model's performance \cite{maldonado2009wrapper}. They often involve a search process, like forward selection or backward elimination, to find the optimal feature subset that maximizes model performance. However, they can be computationally intensive due to the need to train models multiple times for different feature subsets. 
Filter methods, such as Pearson correlation coefcient \cite{liu2020daily}, mutual information \cite{yu2003feature} do not need to perform model training, as a result, the selection of such methods are not refined enough. Filter methods are generally faster but not refined enough than wrapper methods.
Embedded Methods integrate feature selection as part of the model training process \cite{fonti2017feature, ma2021towards, ke2017lightgbm}. Algorithms like LASSO \cite{fonti2017feature} and decision trees \cite{ke2017lightgbm} are typical examples. They are efficient as they perform feature selection and model training simultaneously.

\subsection{Feature Selection for DRS}

\subsubsection{Early Feature Selection} 
Early holistic feature selection in deep recommender systems usually find a subset of k feature fields from the entire set  that significantly enhances the model's predictive accuracy. The selected subset should perform as well as or better than the complete set in terms of the model's performance. Then the final deep recommendation model is retrained with only selected feature fields. Typical representatives include AutoField \cite{wang2022autofield} and AutoFSS \cite{wei2023automatic}. AutoField designs a differentiable controller network that scores each feature field, enabling automatic adjustment in the probability of selecting specific feature fields. On the other hand, AutoFSS initially embeds the search space into a weight-sharing Supernet. It then employs a well-crafted sampling method, which takes into account feature convergence fairness, to train the Supernet. 
SFS\cite{wang2023SFS} scores feature fields by quantifying the difference in loss value when dropping the feature field.
SHARK\cite{zhang2023shark} scores feature fields by quantifying model performance degradation when replacing original feature values and employs direct elimination of low-scoring feature fields.
LPFS\cite{guo2022lpfs} employs smoothed-$l_0$-like function to select a more informative subset of feature fields.
It's noteworthy that OptFS \cite{lyu2023optimizing} adopts a distinct approach compared to other methods for feature field selection. Instead of directly filtering feature fields, OptFS initially sets a learnable gate to identify an optimizing set of features and then retrains the model with this gate fixed. However, due to the requirement of performing a dot product between the gate and the embeddings, this method does not reduce the activated parameter of the embedding layer during training.  It's also important to note that OptFS does not employ an adaptive approach. 
When new feature IDs emerge or when there is a shift in data distribution, OptFS requires a fresh search for the optimal feature subset and subsequent retraining. This limitation highlights the need for more dynamic and adaptable feature selection methods in the ever-evolving landscape of recommender systems.

\subsubsection{Late Feature Selection} 
Late adaptive feature selection in deep recommender systems typically employs a controller network to compute the importance score of each feature field post-embedding layer. Notable examples include AdaFS \cite{lin2022adafs} and MvFS \cite{lee2023mvfs}. AdaFS pioneered this concept and achieved significantly better performance compared to prior holistic feature selection methods, due to its ability to adaptively select features for each user-item interaction. The subsequent work, MvFS, utilizes a multi-view network composed of multiple sub-networks, further enhancing the effectiveness in selecting informative features for individual instances. However, as feature selection occurs after the embedding layer in these methods, they do not reduce the parameter count or the number of embedding lookups of the embedding layer.
\model\xspace is different from previous methods by accomplishing adaptive early feature selection.

\section{Proposed Method}\label{sec:method}
\begin{figure*}[t]
\centering
\includegraphics[width=0.98\textwidth]{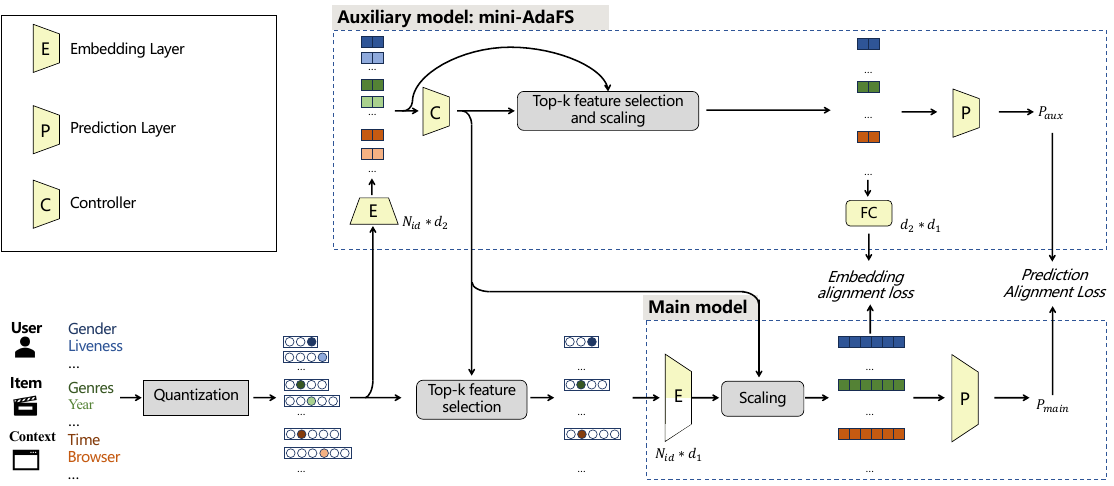}
\caption{\textbf{Conceptual diagram of the proposed adaptive early feature selection (\model) method}. The main model (with an embedding size of $d_1=32$) is responsible for recommendation prediction.
In order to enable \emph{adaptive} feature selection before the embedding layer of the main model, we propose to use an auxiliary model (here a mini-AdaFS with a much smaller embedding size of $d_2=4$) to estimate the importance of the individual features and accordingly select the  $k$ most important features. Besides the regular BCE loss for supervised training, we train the main and auxiliary models in a collaborative manner such that their feature embeddings and the predicted scores are aligned. AEFS maintains the accuracy of AdaFS, whilst saving up to 37.5\% of the embedding parameters activated for forward and backward computation.
}
\label{Figure:overall_framework}
\end{figure*}


\subsection{Overall Architecture}

An overview of the \model framework is illustrated in Fig. \ref{Figure:overall_framework}. 
The framework encompasses two pivotal components: the \auxModel for feature selection and the \mainModel for making final predictions.

A crucial aim of \model is adaptive early feature selection for each user-item interaction. To achieve this, we introduce the \auxModel, positioned in the upper section of Fig. \ref{Figure:overall_framework}. This model is designed to obtain important feature scores that assist in selecting feature fields at the data instance level for the \mainModel. For enhanced efficiency, the \auxModel is equipped with embedding dimensions significantly smaller than those in the \mainModel.

To align the feature importance scores derived from the \auxModel with the \mainModel, we incorporated two collaborative training losses: \EAL and \PAL.  The \EAL encourages close synchronization of embedding values between the \auxModel and the \mainModel, which is crucial for maintaining consistency in feature selection. The \PAL aligns the final prediction outcomes of the \auxModel with the \mainModel, enabling the \auxModel to learn feature interaction patterns similar to the more complex \mainModel.

As a result, the \auxModel effectively supports the \mainModel in adaptive early feature selection. This capability allows \model to not only maintain high accuracy as late feature selection but also optimize activated parameter efficiency in the embedding layer as early feature selection.

\subsection{Preliminary}

More formally, we assume that each user-item-context triplet is represented by a set of $N$ raw features $\{f_1, \ldots, f_N\}$ from N feature fields.
A neural network for DRS typically consists of a nontrainable quantization layer, an embedding layer and a prediction layer. All of our model modifications will base on the basic DRS. We use $M$ to denote a neural network. Its embedding layer is accessed by $M.E$, and its prediction layer is accessed by $M.P$.

\begin{equation} \label{eq:basicDRS}
\left\{ \begin{array}{ll}
 \{x_1, \ldots, x_N\} & \leftarrow \mbox{quantization}(\{f_1,\ldots,f_N\}), \\
 \{e_1, \ldots, e_N\} & \leftarrow M.E(\{x_1, \ldots, x_N\}),\\
 \hat{y} & \leftarrow M.P(\{e_1, \ldots, e_N\})).
       \end{array} \right.
\end{equation}

\subsubsection{Raw feature quantization}
Raw features of DRS mainly include user profile attributes (like gender, liveness), item characteristics (such as genres, release year), and contextual elements (time of interaction, browser type, etc.) \cite{cheng2016wide}. These input features typically encompass both enumerable categorical variables and continuous numerical variables. Numerical features are often transformed into categorical ones through discretization, which is achieved using predefined rules\cite{lin2022adafs}. For categorical features, field-wise quantization is employed, where each feature value is mapped to a high-dimensional vector, facilitating a more nuanced and detailed representation in the recommender models \cite{DBLP:conf/kdd/HePractical}.

\subsubsection{Embedding layer}

The embedding layer is a basic component widely used in deep recommender systems \cite{lyu2022optembed, song2019autoint, guo2017deepfm, SSCDR, wei2021autoias}, which transformes the high-dimensional sparse one-hot vectors into dense latent space for learning feature representations.
In a production recommender system, the ID type features can reach a scale of billions (e.g., \cite{wang2018billion, eksombatchai2018pixie}), and feature crosses are widely utilized \cite{cheng2016wide}. For instance, as illustrated in \cref{tab:parm_emb_dense}, for widely-used recommender models such as MLP \cite{zhang2016deep}, DeepFM \cite{guo2017deepfm}, and DCN \cite{wang2017dcn} on the Avazu and Criteo datasets, the parameter count of the prediction layer is only about one ten-thousandth to five ten-thousandths of that of the embedding layer. As a result, the embedding layer is typically highly sparse and dominates the parameter space \cite{lian2022persia}.
Let $\boldsymbol{PM_n}$ denote a learnable projection matrix of the field $n$.
After projecting each feature field (that is, the embedding lookup), the data instance $\boldsymbol{X}$ is converted to the embedding vector $\boldsymbol{E}$ as:

\begin{equation}
  \begin{aligned}
     \boldsymbol{E} = & \ M.E(\boldsymbol{X}) \\
                    = & \ [\boldsymbol{PM_1}(\boldsymbol{x_1}), \boldsymbol{PM_2}(\boldsymbol{x_2}), \ldots, \boldsymbol{PM_N}(\boldsymbol{x_N})] \\
                    = & \ [\boldsymbol{e_1}, \boldsymbol{e_2}, \ldots, \boldsymbol{e_N}],
  \end{aligned}
\end{equation}
where $\boldsymbol{e_n}$ = $\boldsymbol{PM_n(x_n)}$ denotes the embedding of each field.

\subsubsection{\textbf{Prediction Layer}}   This layer takes \xr{the} feature embeddings $ \boldsymbol{E}$ as input, and \xr{outputs a probabilistic score $\hat{y}$, indicating the likelihood of an event like a click or a transaction}:
 \begin{equation}
\hat{y}= M.P(\boldsymbol{E}).
 \end{equation}
\xr{A typical approach to implementing this layer} is the Multiple Layer Perceptron (MLP). \xr{There are also more sophisticated implementations available}, such as the Deep Factorization Machine (DeepFM) \cite{guo2017deepfm} and the Deep \& Cross Network (DCN) \cite{wang2017dcn}.
Normally, a binary cross-entropy (BCE) loss is used for supervised training:
 \begin{equation}
\mathcal{L}_{BCE} =-y \log (\hat{y})-(1-y) \log (1-\hat{y}),
 \end{equation}
where $y \in\{0,1\}$ is a \xr{ground-truth} label, indicating whether an action like a click or a transaction has truly occurred.

\subsection{AdaFS in a Nutshell}
To selectively exploit informative features for each data point, the adaptive feature selection \cite{lin2022adafs} method uses a controller network $M.C$ to compute the importance score for each feature embedding:
\begin{equation}
 \boldsymbol{S} = M.C(\boldsymbol{E}) = M.C([ \boldsymbol{e_1}, ..., \boldsymbol{e_N} ] ).
\end{equation}
The Controller comprises three key operations: Batch Normalization, a Fully-Connected Layer mapping, and the application of the Softmax function. This importance score is then used to generate a weighted feature vector by multiplying it with the corresponding feature embedding.
The weighted feature vector $\boldsymbol{E^{s}}$ is obtained as follows:
\begin{equation}
 \boldsymbol{E^{s}} = Scaling(\boldsymbol{E}, \boldsymbol{S}) = [s_1\boldsymbol{e_1}, s_2\boldsymbol{e_2}, ..., s_N\boldsymbol{e_N}],
\end{equation}
where $s_n$ represents importance score of the $n$-th field computed by the controller.
$s_n$ can be either a real value (soft selection) or a binary value (hard selection) \cite{lin2022adafs}.  Soft selection scales feature values, while hard selection can eliminate features. However, hard selection does not reduce the computational load of the embedding layer, as it still requires the initial acquisition of the feature embeddings.

\subsection{Adaptive Early Feature Selection}

\xr{The proposed adaptive early feature selection (AEFS) method is to combine the precision of adaptive late FS with the efficiency of early FS. In particular}, \model has two key components: the \mainModel (with an embedding size of $d_1$) and the \auxModel (here a mini-AdaFS with a much smaller embedding size of $d_2$). The \auxModel, can be instantiated either as a {mini-AdaFS} or as a \xr{mini-MvFS, performs adaptive feature selection before the embedding layer of the \mainModel}. Later we shall use $M_m$ to indicate the \mainModel and $M_a$ to indicate the \auxModel.

\subsubsection{MiniAdaFS as the \AuxModel}
The \auxModel is engineered to assist in the adaptive early feature selection for the \mainModel. It is characterized by a reduced-size embedding layer $M_a.E$ and a  prediction layer $M_a.P$. The embedding layer of the \auxModel transforms data instance $\boldsymbol{X}$ into auxiliary embeddings $\boldsymbol{E_{a}}$, as shown below:

\begin{equation}
    \boldsymbol{E_{a}} = M_{a}.E(\boldsymbol{X}) =  [\boldsymbol{e_{a1}}, \boldsymbol{e_{a2}}, \ldots, \boldsymbol{e_{aN}}].
\end{equation}

The importance scores for each feature are computed using a controller network, $ M_{a}.C $, taking $ \boldsymbol{E_{a}} $ as input:
\begin{equation}
    \boldsymbol{S} = M_{a}.C(\boldsymbol{E_{a}}) =  [s_1, s_2, \ldots, s_N].
\end{equation}

To get the selected features, we first perform K-Max Pooling to select the indices of the top $k$ features:

\begin{equation}
    \boldsymbol{I} = \text{k\_max\_pooling}(\boldsymbol{S}, k).
\end{equation}
The selected top $k$ importance weights is 
\begin{equation}
    \boldsymbol{W} = \text{L1norm}(\text{Selection}(\boldsymbol{S}, \boldsymbol{I})).
\end{equation}
Subsequently, the top $k$ features are selected and scaled, forming a subset of auxiliary embeddings $\boldsymbol{E^{s}{a}}$:
\begin{equation}
\begin{aligned}
    \boldsymbol{E^{s}_{a}} &= \text{Scaling}(\text{Selection}(\boldsymbol{E_{a}}, \boldsymbol{I}), \boldsymbol{W}) \\
    &= [\boldsymbol{e_{I_1}} \cdot w_1, \boldsymbol{e_{I_2}} \cdot w_2, \ldots, \boldsymbol{e_{I_k}} \cdot w_k].
\end{aligned}
\end{equation}
The \emph{Scaling} step adjusts the magnitude of each selected embedding using the normalized importance weights so that they add up to one. Specifically, it performs element-wise multiplication between each selected embedding and its corresponding weight. 
This mechanism ensures that the selected features contribute proportionally to their importance in the final representation. 
The final prediction score from the \auxModel, $ P_{a} $, is obtained using a prediction layer:
\begin{equation}
    P_{a} = M_a.P(\boldsymbol{E^{s}_{a}}).
\end{equation}

\subsubsection{Adaptive Early Feature Selection for the  \MainModel}
The \mainModel uses the top-k index $\boldsymbol{I}$ from the \auxModel for early feature selection. The data instance with the selected feature fields for the \mainModel, indicated by $\boldsymbol{X^{s}}$, are determined as follows:
\begin{equation}
    \boldsymbol{X^{s}}= \text{Selection}(\boldsymbol{X}, \boldsymbol{I}).
\end{equation}

The main embeddings for the selected feature fields are produced by the main embedding layer. 
After a scaling operation, we obtain the output of the main embedding layer $\boldsymbol{E^{s}_{m}}$ as
:
\begin{equation}
\begin{aligned}
    \boldsymbol{E^{s}_{m}} &=\text {Scaling} ( M_{m}.E(\boldsymbol{X^{s}}), W ) \\
                              &= [\boldsymbol{e_{m1}}, \boldsymbol{e_{m2}}, ..., \boldsymbol{e_{mk}}]
\end{aligned}
\end{equation}

Finally, the prediction score for the \mainModel $ P_{m} $ is:
\begin{equation}
    P_{m} = M_{m}.P(\boldsymbol{E^{s}_{m}}).
\end{equation}

To sum up, \xr{the auxiliary model helps the main model to} efficiently select the most pertinent features per each user-item interaction. \xr{More specifically, the auxiliary model}'s capability to generate importance scores directs the \mainModel's feature selection, ensuring only the most informative features to be further processed.

\subsection{Multi-Scale Collaborative Training}

\xr{While the idea of using a mini-AdaFS seems simple, how to properly train such an auxiliary model that can effectively select features for the main model is non-trivial. The two models need to be collaboratively trained. To that end, we develop a}
multi-scale collaborative training strategy.
Aligning the models is akin to the teacher-student method in knowledge distillation but with a crucial difference. Unlike knowledge distillation, where a simpler model mimics the final outputs of a more complex one, this alignment focuses on synchronizing the feature importance scores between the two models.  Here, the \auxModel is trained not to replicate the entire predictive capability of the \mainModel but to understand and align the prioritization of features.  By doing so, the \auxModel, although less complex, can effectively discern the relevance of different features in a manner consistent with the \mainModel.  This strategy significantly simplifies the learning objective and shifts its focus from emulating detailed outputs to understanding feature significance.  This approach not only maintains the efficiency of the recommendation process but also ensures the relevance and accuracy of the features used, ultimately contributing to more sophisticated and computationally intensive recommender systems.

\subsubsection{Fine-grained embedding alignment}
To synchronize embeddings between the \auxModel and the \mainModel, we introduce an embedding alignment loss \( \mathcal{L}_{EA} \) as

\begin{equation}
    \mathcal{L}_{EA} = \frac{1}{M} \sum_{i=1}^{M} \left( M_a.FC(\boldsymbol{E_{a}^{s\ \ (i)}}) - \boldsymbol{E_{m}^{s\ \ (i)}} \right)^2,
\end{equation}
where \( M \) denotes the number of instances in a mini-batch. The selected auxiliary embeddings \( \boldsymbol{E_{a}^{s}} \) are mapped to the same dimension as the main embeddings \( \boldsymbol{E_{m}^{s}} \) through a fully connected layer ($M_a.FC$). This fine-grained alignment ensures that the embeddings from the \auxModel closely align with those of the \mainModel, promoting consistency in feature importance assessment and enhancing feature selection quality in recommender systems.

\subsubsection{Coarse-grained prediction alignment}

To ensure that the final prediction outcomes of the \auxModel align with those of the \mainModel, we introduce an prediction alignment loss $\mathcal{L}_{PA}$ as
\begin{equation}
    \mathcal{L}_{PA} = \frac{1}{M} \sum_{i=1}^{M} (P_{a}^{(i)} - P_{m}^{(i)})^2,
\end{equation}
where \( P_{a} \) represents the prediction score from the \auxModel, and \( P_{m} \) represents the prediction score from the \mainModel. The objective of $\mathcal{L}_{PA}$ is to minimize the squared differences between the predictions of the two models across all instances in the mini-batch. By reducing this discrepancy, the \auxModel learns to approximate the prediction behavior of the \mainModel. This alignment process helps the \auxModel to understand the underlying patterns and interactions that drive the predictions of the \mainModel, thereby adopting a similar decision-making process.

\subsubsection{Optimization}
We train AEFS model to predict interactions of user-item pairs.
The loss function is defined as follows:
\begin{equation}
 \begin{aligned}
    \min_{\boldsymbol{\theta_{{m}}}, \boldsymbol{\theta_{{a}}}} \frac{1}{M}\sum_{i=1}^{M} \mathcal{L}_{BCE} + \mathcal{L}_{PA} + \mathcal{L}_{EA},
 \end{aligned}
\end{equation}
\noindent
where $\boldsymbol{\theta_{{a}}}$, $\boldsymbol{\theta_{{m}}}$  denote the parameters of the \mainModel and \auxModel, including the embedding components and the subsequent layers.  $M$ is the instance number in a mini-batch, indexed by $(i)$. The training algorithm is shown in Alg. \ref{alg:AEFS}.

\begin{algorithm}
\caption{AEFS in a PyTorch style}
\label{alg:AEFS}

\begin{PythonB}
Input: Training data loader D={(X, y)};
~~~~~~~Number of feature fields N;
~~~~~~~Feature selection rate r (default: 0.5);
~~~~~~~Main embedding size d_1 (default: 32);
~~~~~~~Auxiliary embedding size d_2 (default: 4);
Output: Trained main model M_m
~~~~~~~~Trained auxiliary model M_a

k = N*r
optimizer = torch.optim.Adam({M_m.params, M_a.params})
for i=1, 2, ..., MAX_EPOCHS:
    for mini-batch (X, y) in D:
        optimizer.zero_grad()

        e_a = M_a.E(X, d_2)
        s = M_a.C(e_a)
        index = k_max_pooling(s, k)
        w = l1_norm(s[index])
        e_a = scaling(e_a[index], w)
        p_a = M_a.P(e_a)

        e_m = M_m.E(X[index], d_1)
        e_m = scaling(e_m, w)
        p_m = M_m.P(e_m)

        loss_bce = BCE(p_a, y) + BCE(p_m, y)
        loss_ea = MSE(e_m, M_a.FC(e_a))
        loss_pa = MSE(p_m, p_a)
        loss = loss_bce + loss_ea + loss_pa
        loss.backward()
        optimizer.step()
\end{PythonB}

\end{algorithm}

\subsection{Activated Parameter Efficiency} \label{sec:ParameterEff}


We measure the efficiency of \model with two criteria, 
\emph{i.e.} the reduction in overall activated embedding parameters and the decrease in embedding lookups by the \mainModel. 
Suppose 50\% of the feature fields are selected for the \mainModel.

\textbf{Reduction in Activated Embedding Parameters}. The overall reduction in activated embedding parameters is calculated as the difference between the decrease in parameters of the \mainModel and the increase in parameters of the \auxModel. Suppose the number of distinct feature IDs is \( N_{id} \). 
The ratio between the embedding sizes of \mainModel and \auxModel is \( d_1 = 8d_2 \).
The percentage reduction in overall activated embedding parameters, denoted as \( \Delta PaE \), can be quantified as:
\begin{equation}
\resizebox{.8\hsize}{!}{$
 \Delta PaE = \frac{N_{id} \times d_1 \times 50\% - N_{id} \times d_2}{N_{id} \times d_1} = 50\% - \frac{d_2}{d_1} = 37.5\%
 $}
\end{equation}

\textbf{Reduction in Embedding Lookups}. The reduction in embedding lookups by the \mainModel, denoted as \( \Delta EL \), is directly proportional to the early feature selection ratio. With a 50\% early feature selection, the reduction is 50\%.

\section{Experiments}

For a comprehensive evaluation of the proposed \model,
we design experiments to answer the following research questions:
\begin{itemize}
    \item \textbf{RQ1}: Could \model achieve comparable performance to existing adaptive feature selection methods while significantly reducing the activated parameters of the embedding layer?

    \item \textbf{RQ2}: How transferable is \model? Does altering the \auxModel significantly impact the overall performance?  

    \item \textbf{RQ3}: How does $\Delta PaE$ affect the inference and training efficiency of \model?  
    
    \item \textbf{RQ4}: How does \model compare to AdaFS in terms of activated parameters \xr{for model} inference?  
    \item \textbf{RQ5}:  What is the impact of the individual components in \model? 

\end{itemize}

\subsection{Experimental Setup}

\subsubsection{Datasets}
\xr{We adopt three benchmark datasets, \ie Avazu, Criteo and KDD12.}
Following \cite{lin2022adafs, wang2022autofield}, per dataset we adopt an 8:1:1 random split of training, validation, and test sets. \\
$\bullet$ \textbf{Avazu}\footnote{http://www.kaggle.com/c/avazu-ctr-prediction}: Spanning 10 days of click logs, the Avazu dataset comprises approximately 40 million user-click records across 22 feature fields, and some features are anonymous. \\
$\bullet$ \textbf{Criteo}\footnote{https://www.kaggle.com/c/criteo-display-ad-challenge}: This dataset includes a week's worth of ad click data, featuring 45 million user-click records on delivered ads. It contains 26 categorical and 13 numerical feature fields.  In line with established practices \cite{zhu2020fuxictr}, we apply discretization to numerical values, transforming each value \( x \) to \( \lfloor\log^2(x)\rfloor \) when \( x > 2 \); otherwise, we set \( x = 1 \). All features in this dataset are anonymized. \\
$\bullet$ \textbf{KDD12}\footnote{http://www.kddcup2012.org/c/kddcup2012-track2/data}: This dataset consists of training instances extracted from search session logs. It includes 11 categorical fields, and the click field represents the number of times a user clicked on an ad. To handle rare features, we replace infrequent categories with an "OOV" (out-of-vocabulary) token using a minimum frequency threshold of 10.

\subsubsection{Metrics}
Following the previous works~\cite{FM,guo2017deepfm}, we employ widely recognized evaluation metrics for CTR prediction: \textbf{AUC} (Area Under the ROC Curve) and \textbf{Logloss} (cross-entropy). Notably, even a $\mathbf{0.1\%}$ improvement in these metrics can signify considerable advancements in CTR prediction tasks \cite{qu2016ipnn, guo2017deepfm}. To assess the activated parameter efficiency, we introduce \EmbRatio, representing the reduction ratio in activated embedding layer parameters achieved by a specific method.
The \EmbRatio for \model is calculated as the reduction in the \mainModel minus the increase due to \auxModel. 
For the test of significance, we adopt a two-sided t-test
(p-value $<$ 0.05).

\begin{table*}[t]
\centering \setlength{\tabcolsep}{7.0pt}
\caption{\textbf{The overall performance of different feature selection (FS) methods}. Our \model method is as effective as the late FS methods, \ie AdaFS and MvFS (with no statistically significant difference), while using 37. 5\% less activated parameters.
\textbf{Bold} font indicates the best runs.
}
\label{tab:overall_performance}

\begin{tabular}{@{}l|l|rrr|rrr|rrr@{}}
\toprule
\multirow{3}{*}{\textbf{Dataset}} & \multicolumn{1}{c|}{\multirow{3}{*}{\textbf{Methods}}} & \multicolumn{9}{c}{\textbf{Choice of the prediction layer (P) in the main model}} \\
 & \multicolumn{1}{c|}{} & \multicolumn{3}{c|}{{MLP}} & \multicolumn{3}{c|}{{DeepFM}} & \multicolumn{3}{c}{{DCN}} \\
 & \multicolumn{1}{c|}{} & \multicolumn{1}{l}{\textit{AUC$\uparrow$}} & \multicolumn{1}{l}{\textit{Logloss$\downarrow$}} & \multicolumn{1}{l|}{\textit{\EmbRatio $\uparrow$}} & \multicolumn{1}{l}{\textit{AUC$\uparrow$}} & \multicolumn{1}{l}{\textit{Logloss$\downarrow$}} & \multicolumn{1}{l|}{\textit{\EmbRatio $\uparrow$}} & \multicolumn{1}{l}{\textit{AUC$\uparrow$}} & \multicolumn{1}{l}{\textit{Logloss$\downarrow$}} & \multicolumn{1}{l}{\textit{\EmbRatio $\uparrow$}} \\
 \midrule
\multirow{8}{*}{\emph{Avazu}} 
 & No FS & 0.7763 & 0.3823 & 0.0\% & 0.7798 & 0.3797 & 0.0\% & 0.7793 & 0.3801 & 0.0\% \\
 & AutoField \cite{wang2022autofield} & 0.7584 & 0.3908 & \textbf{37.5\%} & 0.7736 & 0.3830 & \textbf{37.5\%} & 0.7519 & 0.3946 & \textbf{37.5\%} \\
 & OptFS \cite{lyu2023optimizing} & 0.7779 & 0.3812 & \textbf{37.5\%} & 0.7737 & 0.3830 & \textbf{37.5\%} & 0.7711 & 0.3846 & \textbf{37.5\%} \\
 & LPFS \cite{guo2022lpfs} & 0.7749 & 0.3828 & \textbf{37.5\%} & 0.7759 & 0.3786 & \textbf{37.5\%} & 0.7784 & 0.3803 & \textbf{37.5\%} \\
 & SHARK \cite{zhang2023shark} & 0.7764 & 0.3823 & \textbf{37.5\%} & 0.7786 & 0.3810 & \textbf{37.5\%} & 0.7778 & 0.3813 & \textbf{37.5\%} \\
 & AdaFS \cite{lin2022adafs} & 0.7785 & 0.3808 & 0.0\% & 0.7791 & 0.3808 & 0.0\% & 0.7794 & 0.3800 & 0.0\% \\
 & MvFS \cite{lee2023mvfs} & \textbf{0.7796} & 0.3801 & 0.0\% & \textbf{0.7816} & \textbf{0.3786} & 0.0\% & 0.7812 & 0.3789 & 0.0\% \\
 & \model~(miniAdaFS) & 0.7785 & 0.3808 & \textbf{37.5\%} & 0.7793 & 0.3802 & \textbf{37.5\%} & 0.7794 & 0.3802 & \textbf{37.5\%} \\
 & \model~(miniMvFS) & 0.7794 & \textbf{0.3800} & \textbf{37.5\%} & 0.7814 & 0.3789 & \textbf{37.5\%} & \textbf{0.7814} & \textbf{0.3788} & \textbf{37.5\%} \\ \midrule
 
\multirow{8}{*}{\emph{Criteo}} & No FS & 0.8025 & 0.4496 & 0.0\% & 0.8066 & 0.4452 & 0.0\% & 0.8053 & 0.4464 & 0.0\% \\
 & AutoField & 0.7963 & 0.4546 & \textbf{37.5\%} & 0.8012 & 0.4499 & \textbf{37.5\%} & 0.7961 & 0.4541 & \textbf{37.5\%} \\
 & OptFS & 0.8001 & 0.4512 & \textbf{37.5\%} & 0.8060 & 0.4458 & \textbf{37.5\%} & 0.8024 & 0.4488 & \textbf{37.5\%} \\
 & LPFS & 0.8004 & 0.4542 & \textbf{37.5\%} & 0.8061 & 0.4455 & \textbf{37.5\%} & 0.8035 & 0.4479 & \textbf{37.5\%} \\
 & SHARK & 0.8028 & 0.4489 & \textbf{37.5\%} & 0.8035 & 0.4481 & \textbf{37.5\%} & 0.8032 & 0.4485 & \textbf{37.5\%} \\
 & AdaFS & 0.8056 & 0.4467 & 0.0\% & 0.8057 & 0.4463 & 0.0\% & 0.8064 & 0.4468 & 0.0\% \\
 & MvFS & 0.8055 & 0.4467 & 0.0\% & 0.8067 & 0.4448 & 0.0\% & \textbf{0.8066} & \textbf{0.4449} & 0.0\% \\
 & \model~(miniAdaFS) & \textbf{0.8057} & \textbf{0.4465} & \textbf{37.5\%} & 0.8060 & 0.4460 & \textbf{37.5\%} & 0.8062 & 0.4471 & \textbf{37.5\%} \\
 & \model~(miniMvFS) & 0.8054 & 0.4484 & \textbf{37.5\%} & \textbf{0.8067} & \textbf{0.4447} & \textbf{37.5\%} & \textbf{0.8066} & \textbf{0.4449} & \textbf{37.5\%} \\ \midrule

\multirow{8}{*}{\emph{KDD12}} & No FS & 0.7774 & 0.1575 & 0.0\% & 0.7691 & 0.1582 & 0.0\% & 0.7787 & 0.1574 & 0.0\% \\
 & AutoField & 0.7642 & 0.1605 & \textbf{37.5\%} & 0.7648 & 0.1587 & \textbf{37.5\%} & 0.7674 & 0.1586 & \textbf{37.5\%} \\
 & OptFS & 0.7702 & 0.1592 & \textbf{37.5\%} & 0.7720 & 0.1582 & \textbf{37.5\%} & 0.7752 & 0.1579 & \textbf{37.5\%} \\
 & LPFS & 0.7739 & 0.1587 & \textbf{37.5\%} & 0.7753 & 0.1572 & \textbf{37.5\%} & 0.7749 & 0.1573 & \textbf{37.5\%} \\
 & SHARK & 0.7745 & 0.1581 & \textbf{37.5\%} & 0.7751 & 0.1574 & \textbf{37.5\%} & 0.7760 & 0.1571 & \textbf{37.5\%} \\
 & AdaFS & 0.7750 & 0.1587 & 0.0\% & 0.7740 & 0.1577 & 0.0\% & 0.7772 & 0.1572 & 0.0\% \\
  & MvFS & 0.7791 & 0.1572 & 0.0\% & \textbf{0.7785} & \textbf{0.1567} & 0.0\% & 0.7796 & 0.1563 & 0.0\% \\
 & \model (miniAdaFS) & 0.7755 & 0.1590 & \textbf{37.5\%} & 0.7755 & 0.1575 & \textbf{37.5\%} & 0.7781 & 0.1574 & \textbf{37.5\%} \\
 & \model (miniMvFS) & \textbf{0.7795} & \textbf{0.1570} & \textbf{37.5\%} & \textbf{0.7785} & \textbf{0.1567} & \textbf{37.5\%} & \textbf{0.7798} & \textbf{0.1562} & \textbf{37.5\%}
 \\

 \bottomrule
\end{tabular}

\end{table*}

\subsubsection{Baseline Methods and Backbone Models}
We select feature field selection baselines to study based on their representatives and their availability. The baselines that satisfy the following three criteria are chosen: 1) The source code is publicly available.  2) The granularity of  selection is feature field. 3) The paper is peer reviewed if it is proposed in a research paper.
As a result, we select four feature selection baselines:
(i) \textbf{AutoField}\footnote{https://github.com/Dave-AdamsWANG/AutoField} \cite{wang2022autofield}, which leverages neural architecture search techniques \cite{liu2018darts} for field-level informative feature selection. We drop 37.5\% feature fields for fair comparison; (ii) \textbf{OptFS}\footnote{https://github.com/fuyuanlyu/optfs} \cite{lyu2023optimizing}, adopting feature value level search; 
(iii) \textbf{LPFS} \cite{guo2022lpfs}, which uses a smoothed-$l_0$ function to select informative feature fields;  
(iv) \textbf{SHARK} \cite{zhang2023shark}, which scores feature fields based on performance degradation and eliminates low-scoring ones. 
We use the implementations of LPFS and SHARK provided in ERASE \cite{jia2024erase}.
(v)  \textbf{AdaFS}\footnote{https://github.com/Applied-Machine-Learning-Lab/AdaFS} \cite{lin2022adafs}, employing a controller network to select the most relevant features for each sample; (vi) \textbf{MvFS}\footnote{https://github.com/dudwns511/MvFS\_CIKM23} \cite{lee2023mvfs}, an extension of AdaFS that utilizes a multi-view network with multiple sub-networks, each dedicated to assessing feature importance for distinct data segments with varying feature patterns. These baselines are applied to mainstream backbone models including MLP \cite{zhang2016deep}, DeepFM \cite{guo2017deepfm}, and DCN \cite{wang2017dcn}.

\subsubsection{Implementation Details}
Main hyper-parameters for \model are empirically set as follows.
The embedding dimensionalities $d_1$ and $d_2$ are set to 32 and 4, respectively. The size of a mini-batch is 2,048.
Following previous work \cite{lin2022adafs}, Adam optimizer, Batch Normalization, and Xavier initialization \cite{Xavier} are adopted.
Our implementation is based on a public PyTorch library for CTR prediction\footnote{https://github.com/rixwew/pytorch-fm}.
For other baseline methods, we refer to the official implementation. Unless otherwise specified, all experiments in this study by default utilize the AdaFS framework and an MLP to implement the prediction layer. 


\subsection{Results}

\subsubsection{Overall Performance (RQ1)}
The overall performance of our different feature selection methods, across three different backbone models on three benchmark datasets, is reported in \cref{tab:overall_performance}. It is important to note that for our \model, the embedding size of \auxModel is only one-eighth of that of the \mainModel. Our observations are summarized below:

Firstly,
our \model demonstrates effectiveness in comparison with other baseline methods.
In terms of effectiveness, AutoField, which removes 37.5\% of features, shows a significant decrease in AUC and Logloss across three datasets and all backbone model scenarios, compared to the No Feature Selection (No FS) approach. This aligns with previous replication studies of AutoField \cite{lin2022adafs, lee2023mvfs, lyu2023optimizing}.
OptFS \cite{lyu2023optimizing}, LPFS \cite{guo2022lpfs}, and SHARK \cite{zhang2023shark} generally underperform relative to adaptive selection methods, and in some cases, even fall short of the No FS baseline.
As it requires a fresh search for optimal feature subsets and retraining when new feature IDs emerge or data distribution shifts, we consider such methods to be suboptimal. 
The adaptive selection methods AdaFS and MvFS maintain or significantly improve performance compared to the No FS approach. \model~(miniAdaFS) and \model~(miniMvFS) exhibit similar AUC metrics to AdaFS and MvFS, with no significant differences, indicating that the \auxModel could effectively select features and enhances performance, even with only one-eighth the embedding parameters of the \mainModel.

Secondly, our \model shows efficiency compared to AdaFS and MvFS. 
The adaptive selection methods AdaFS and MvFS, which require initial acquisition of feature embeddings before assessing importance, show no change in \EmbRatio. \model, however, first generates important scores through the \auxModel and then selects features for the \mainModel at the data instance level, resulting in an \EmbRatio of 37.5\%. This approach simultaneously achieves adaptive feature selection and a reduction in \EmbRatio.

\begin{table}[t]
\centering \setlength{\tabcolsep}{5.0pt}
\caption{\textbf{Evaluating the transferable performance of \model}, {when the prediction layer (P) of the auxiliary model differs from its counterpart in the main model}. No statistically significant difference is observed as determined by a two-sided t-test (p-value $<0.05$). }
\label{tab:transfer}

\begin{tabular}{@{}llrrrr@{}}
\toprule
\multicolumn{2}{c}{\textbf{Choice of P}} &
\multicolumn{2}{c}{\textbf{Avazu}} & \multicolumn{2}{c}{\textbf{Criteo}} \\
\textit{Main} & \textit{Auxiliary} & \multicolumn{1}{l}{\textit{AUC}$\uparrow$} & \multicolumn{1}{l}{\textit{Logloss}$\downarrow$} & \multicolumn{1}{l}{\textit{AUC}$\uparrow$} & \multicolumn{1}{l}{\textit{Logloss}$\downarrow$} \\ \midrule
\multirow{3}{*}{MLP} & MLP & 0.7785 & \textbf{0.3808} & \textbf{0.8060} & \textbf{0.4466} \\
 & DeepFM & \textbf{0.7787} & 0.3813 & 0.8057 & 0.4468 \\
 & DCN & 0.7783 & 0.3810 & 0.8057 & 0.4470 \\ \midrule
\multirow{3}{*}{DeepFM} & MLP & 0.7793 & 0.3802 & \textbf{0.8062} & \textbf{0.4458} \\
 & DeepFM & \textbf{0.7796} & \textbf{0.3800} & 0.8060 & 0.4460 \\
 & DCN & 0.7793 & 0.3803 & 0.8060 & 0.4460 \\ \midrule
\multirow{3}{*}{DCN} & MLP & 0.7792 & 0.3804 & 0.8061 & 0.4473 \\
 & DeepFM & 0.7792 & 0.3804 & 0.8062 & 0.4477 \\
 & DCN & \textbf{0.7794} & \textbf{0.3802} & \textbf{0.8062} & \textbf{0.4471} \\
 \bottomrule
\end{tabular}

\end{table}

\begin{figure}[ht]
\centering
    \subfloat[ Inference Efficiency. \label{fig:param_avazu}]{\includegraphics[width=0.8\columnwidth]{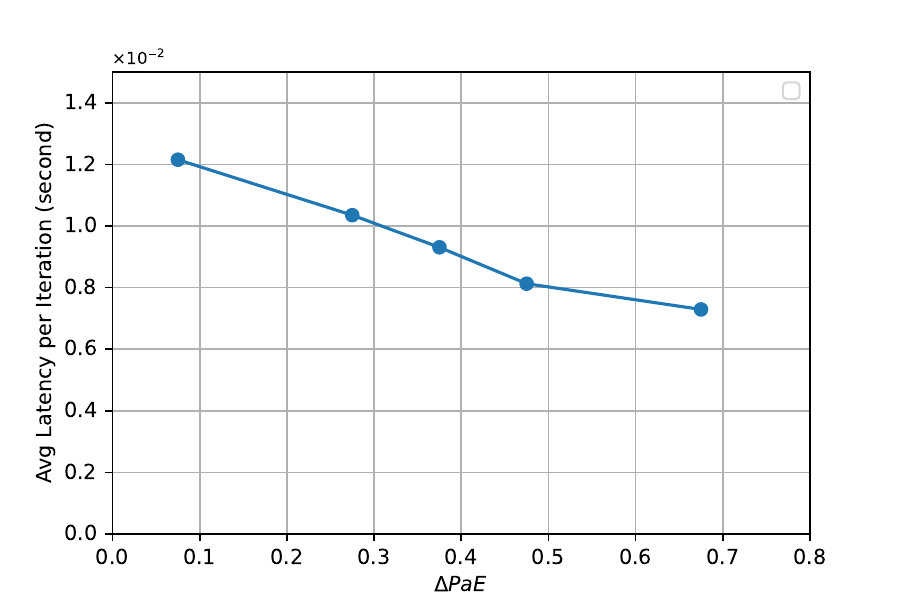}}                   
    
    \subfloat[Training Efficiency. \label{fig:inf_effcience}]{\includegraphics[width=0.8\columnwidth]{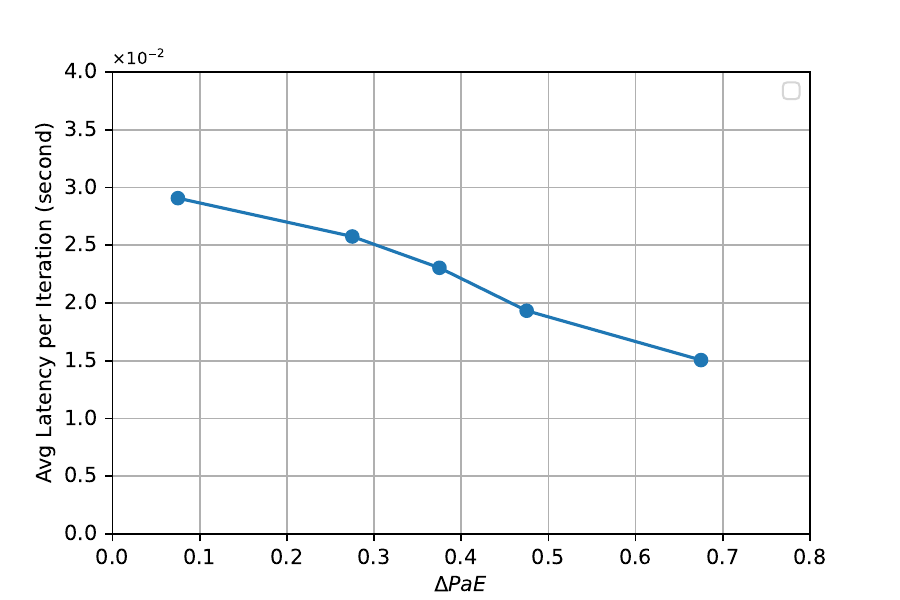}}
    \caption{\textbf{Training and inference efficiency analysis.} We report the average per-sample inference and training time as $\Delta PaE$ increases, using the AEFS model under the Avazu configuration.}
    \label{Figure:training_inf_effcience}
\end{figure}

\subsubsection{Transferability Study (RQ2)}
To investigate the transferability of \model, we fixed the \mainModel and varied the \auxModel to examine its impact on AUC and Logloss performance. We explored the transferability across MLP, DeepFM, and DCN prediction layers on both Criteo and Avazu datasets. The results are presented in \cref{tab:transfer}. It is apparent that all transformations have an effect on performance within a margin of 0.001. Additionally, a two-sided t-test (p-value $<$ 0.05) revealed no statistically significant differences. This suggests that the prediction layer of the \auxModel can be different from the \mainModel without compromising effectiveness. In practical applications, the \mainModel may undergo frequent iterations, but the \auxModel does not necessitate concurrent changes. This aspect underscores the potential of deploying \model in real-world recommender systems.

\subsubsection{Inference and training efficiency with $\Delta PaE$ (RQ3)}
 As discussed in \cref{sec:ParameterEff}, reducing the number of activated embedding parameters lowers the computational cost of embedding lookup operations, which theoretically enhances inference efficiency. 
 We evaluated the impact of $\Delta PaE$ on inference and training efficiency using the AEFS model under the Avazu configuration (vocabulary size = 2,018,012; mini embedding size = 4; main embedding size = 32).
 The results are presented in Fig. \ref{Figure:training_inf_effcience}. Experimental findings indicate that a smaller number of activated embedding parameters leads to reduced inference and training latency. It is worth noting that realizing the full efficiency gains in practice requires system-level design and optimization, which remains an active area of our ongoing research.

\subsubsection{Activated Parameter efficiency Evaluation (RQ4)}
\begin{figure}[t]
\centering
    \subfloat[Avazu. \label{fig:param_avazu}]{\includegraphics[width=0.9\columnwidth]{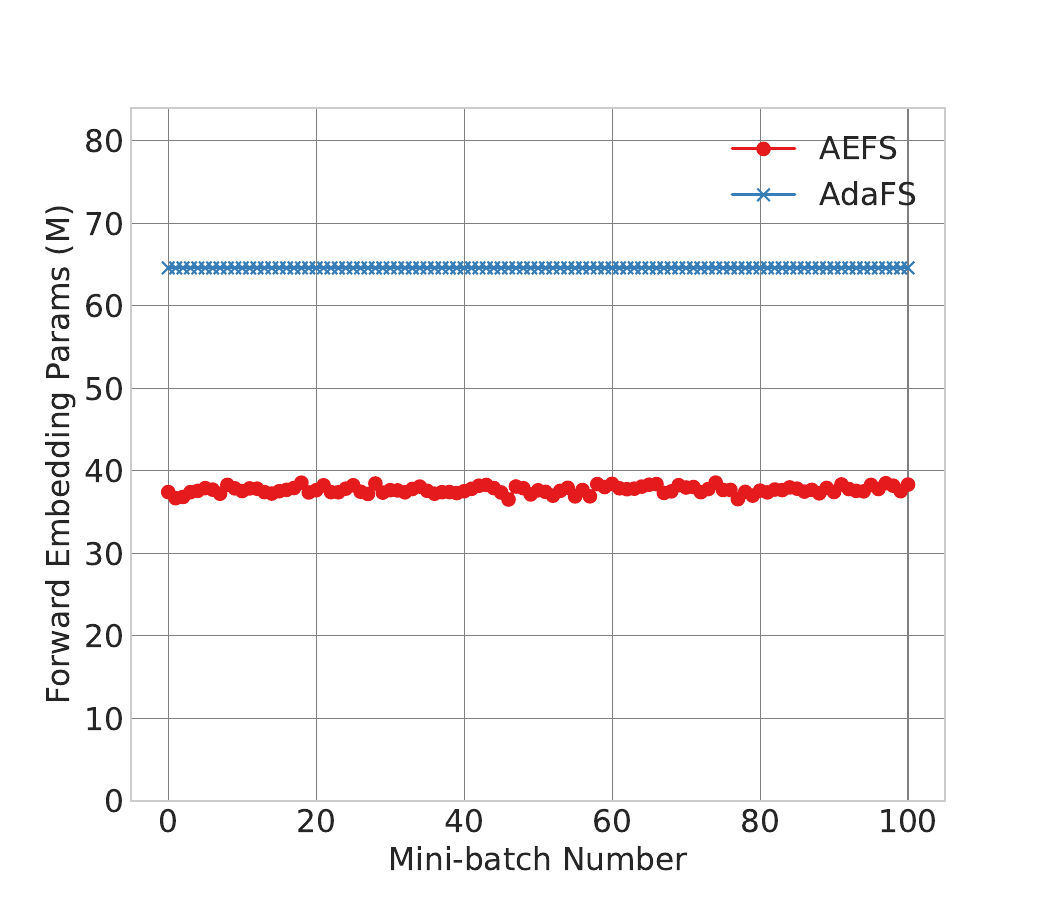}}

    \subfloat[Criteo. \label{fig:param_criteo}]{\includegraphics[width=0.9\columnwidth]{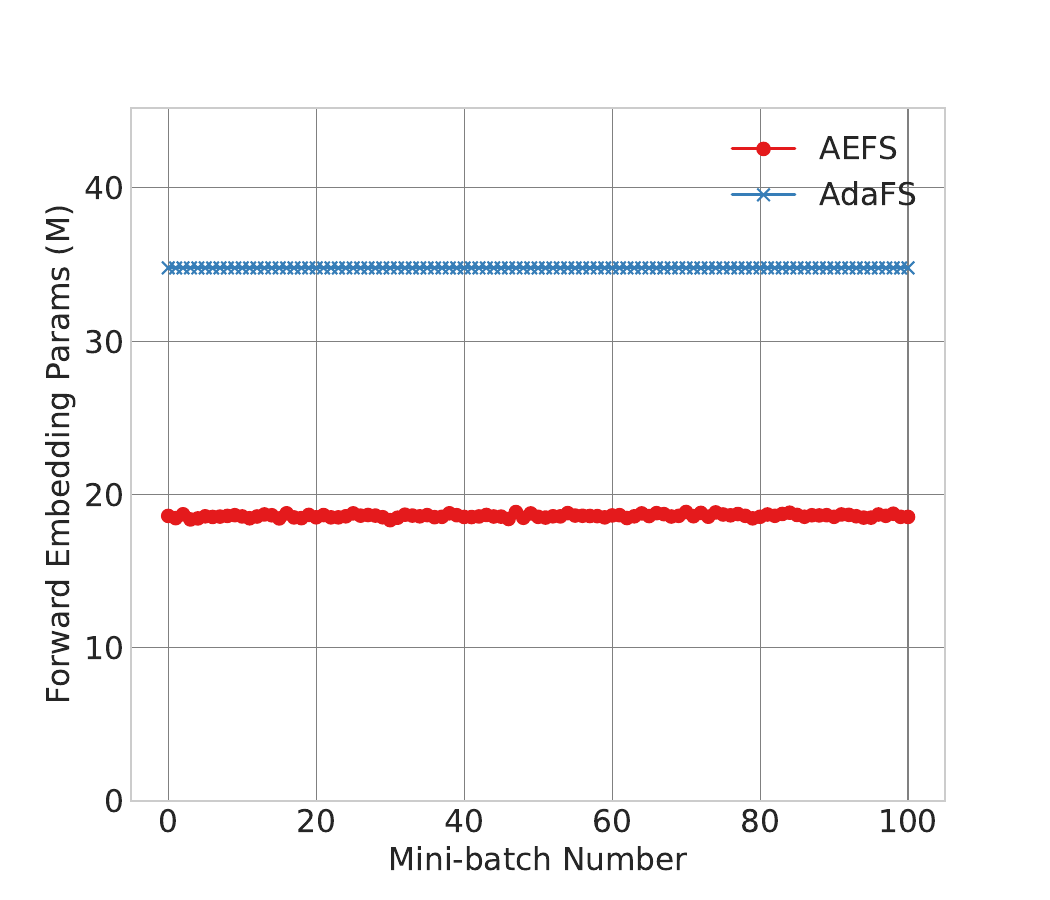}}
    \caption{ \textbf{Comparison of activated embedding layer parameters in Avazu and Criteo.} These figures present the average embedding layer parameters utilized in 100 randomly selected mini-batches, highlighting the differences in parameter usage between the \model and AdaFS methods.
    }
    \label{Figure:param_compare}
\end{figure}

To compare the differences in the activated embedding layer parameters between \model and the AdaFS baseline, we calculated the average embedding layer parameters utilized in 100 randomly selected mini-batches, as shown in Fig. \ref{Figure:param_compare}. Since different feature fields are removed for each sample, the reduction in parameters varies. We observe that \model significantly reduces the average embedding layer parameters by nearly half compared to AdaFS (37.9M vs 64.58M in Avazu and 18.38M vs 34.78M in Criteo). This reduction is attributed to the use of \auxModel for determining feature importance scores, which enabled us to eliminate 50\% of the features in \mainModel. The additional embedding parameters introduced by \auxModel were 8.07M in Avazu and 4.35M in Criteo, while the reduction in embedding parameters in \mainModel amounted to 34.75M in Avazu and 20.75M in Criteo. Overall, this resulted in a substantial decrease in the total activated embedding parameters.

\begin{table}[t]
\centering \setlength{\tabcolsep}{4.0pt}
\caption{\textbf{The influence of the embedding size $d_2$ of the auxiliary model}.
The symbol $\dagger$ indicates that compared to AdaFS,
no statistically significant AUC and Logloss difference is observed.
MLP is used as the P layer.
}
\label{tab:ablation_embsize}

\begin{tabular}{@{}llrrrrr@{}}
\toprule
\multirow{2}{*}{} & \multicolumn{1}{l}{\multirow{2}{*}{\textbf{$R_{es}$}}} & \multicolumn{2}{c}{\textbf{Avazu}} & \multicolumn{2}{c}{\textbf{Criteo}} & \multicolumn{1}{l}{\multirow{2}{*}{\textit{\EmbRatio $\uparrow$}}} \\
 & \multicolumn{1}{l}{} & \multicolumn{1}{l}{\textit{AUC$\uparrow$}} & \multicolumn{1}{l}{\textit{Logloss$\downarrow$}} & \multicolumn{1}{l}{\textit{AUC$\uparrow$}} & \multicolumn{1}{l}{\textit{Logloss$\downarrow$}} & \multicolumn{1}{l}{} \\ \midrule
AdaFS & \multicolumn{1}{l}{n.a.} & 0.7785 & 0.3808 & 0.8056 & 0.4467 & 0.0\% \\ \midrule
\multirow{4}{*}{\model} & 2/32 & 0.7766 & 0.3823 & 0.8043 & 0.4480 & \textbf{43.8\%} \\
 & 4/32$\dagger$ & 0.7785 & 0.3808 & 0.8057 & 0.4465 & 37.5\% \\
 & 6/32$\dagger$ & 0.7789 & 0.3809 & 0.8059 & 0.4467 & 31.3\% \\
 & 16/32 & \textbf{0.7791} & \textbf{0.3806} & \textbf{0.8062} & \textbf{0.4463} & 0.0\% \\
 \bottomrule
\end{tabular}

\end{table}

\begin{table}[t]
\centering \setlength{\tabcolsep}{4.0pt}
\caption{\textbf{Ablation study on the two alignment losses}, \ie EAL and PAL.
Here $*$ denotes statistically significant difference. }
\label{tab:ablation_align}

\begin{tabular}{@{}lllll@{}}
\toprule
\multirow{2}{*}{\textbf{Configuration}} & \multicolumn{2}{c}{\textbf{Avazu}} & \multicolumn{2}{c}{\textbf{Criteo}} \\
 & \multicolumn{1}{l}{\textit{AUC$\uparrow$}} & \multicolumn{1}{l}{\textit{Logloss$\downarrow$}} & \multicolumn{1}{l}{\textit{AUC$\uparrow$}} & \multicolumn{1}{l}{\textit{Logloss$\downarrow$}} \\ \midrule
\model & \textbf{0.7785$^*$} & \textbf{0.3808$^*$} & \textbf{0.8057$^*$} & \textbf{0.4465$^*$} \\
\emph{w/o} EAL & 0.7780  & 0.3814 & 0.8053 & 0.4474 \\
\emph{w/o} PAL & 0.7778 & 0.3812 & 0.8054 & 0.4686 \\
\emph{w/o} EAL and PAL & 0.7776 & 0.3840 & 0.8051 & 0.4633 \\
 \bottomrule
\end{tabular}

\end{table}

\begin{table}[t]
\centering \setlength{\tabcolsep}{5.0pt}
\caption{\textbf{Evaluation of the effect of pretraining and top-k reweight}. The symbol $\dagger$ indicates that no statistically significant AUC and Logloss difference is observed.} 
\label{tab:ablation_adafs}

\begin{tabular}{@{}lllll@{}}
\toprule
\multirow{2}{*}{\textbf{Configuration}} & \multicolumn{2}{c}{\textbf{Avazu}} & \multicolumn{2}{c}{\textbf{Criteo}} \\
 & \multicolumn{1}{l}{\textit{AUC$\uparrow$}} & \multicolumn{1}{l}{\textit{Logloss$\downarrow$}} & \multicolumn{1}{l}{\textit{AUC$\uparrow$}} & \multicolumn{1}{l}{\textit{Logloss$\downarrow$}} \\ \midrule
\model $\dagger$ & \textbf{0.7785} & \textbf{0.3808} & 0.8057 & \textbf{0.4465} \\
- w/o pretraining $\dagger$ & 0.7782 & 0.3811 & \textbf{0.8058} & 0.4467 \\
- w/o top-k reweight & 0.7778 & 0.3819 & 0.8052 & 0.4487 \\
 \bottomrule
\end{tabular}

\end{table}

\subsubsection{Ablation Study  (RQ5)}
In this subsection, we delve into an ablation study of \model to understand the influence of various components and settings on its performance. This study is crucial for determining the optimal configuration of the \model and understanding the impact of each element. We analyze the results from \cref{tab:ablation_embsize} and \cref{tab:ablation_align} to draw insights.

\textbf{Impact of Embedding Size in \auxModel}.
This part of our study focuses on the impact of the embedding size ratio ($R_{es}$) in the \auxModel within the \model framework. We explore how different ratios of embedding sizes ( 2/32, 4/32, 6/32, and 16/32) influence the performance of \model. As illustrated in \cref{tab:ablation_embsize}, for $R_{es}$=2/32, \model demonstrates a substantial performance reduction compared to AdaFS.  For larger $R_{es}$ values of 4/32, 6/32, the differences between \model and AdaFS are not statistically significant, as confirmed by a two-sided t-test with a p-value of less than 0.05. When \(R_{es}\) increases to 16/32, there is a significant improvement in performance, but the \EmbRatio decreases to 0.0\%, indicating no change in the activated embedding parameters of the model.
This suggests that as the size ratio increases, the model became less efficient in using embedding parameters.  Based on these insights, we recommend an $R_{es}$ of 4/32 as the optimal choice for \model. This particular ratio strikes a balance between maintaining model efficiency and ensuring competitive performance in comparison to AdaFS.

\textbf{Effectiveness of the two alignment losses}.
We further evaluated the importance of \EAL (EAL) and \PAL (PAL) in \model, as shown in \cref{tab:ablation_align}. The experimental setup included three \model variants: \model without EAL, \model without PAL, and \model without both EAL and PAL. The results across both datasets indicate that the inclusion of EAL and PAL contributes to the best performance in terms of AUC and Logloss. Removing either EAL or PAL leads to a noticeable drop in the effectiveness of the model, and removing both results in the most significant performance decrease. This underscores the critical role of these alignment strategies in ensuring the coherence between the \auxModel and the \mainModel, thereby optimizing the feature selection process and enhancing the overall performance of \model.

The ablation study highlights two key findings: Firstly, the embedding size in the \auxModel plays a significant role in balancing performance and efficiency, where an optimal embedding size can lead to reduced parameter usage without significantly compromising the model's accuracy. Secondly, the \EAL and \PAL  are vital components of \model, as they ensure the model's effectiveness by aligning the feature importance scores and predictive capabilities of the auxiliary and main models. These insights are instrumental in guiding future enhancements and implementations of \model in real-world applications.

\textbf{Evaluation of the effect of pretraining and top-k reweight}.
In late selection methods such as adaFS and MvFS, both pretraining and the top-k reweighting technique are commonly applied. To investigate whether these strategies remain effective within the bi-model setting of \model, we conducted ablation studies, with results presented in \cref{tab:ablation_adafs}.

The experimental results indicate that the pretraining strategy has minimal impact on recommendation performance, whereas removing reweighting has a substantial effect on outcomes. This suggests that using the two alignment losses to supervise the auxiliary feature selection model reduces the need for pretraining. For top-k reweighting, adjusting feature weights to sum to one after selection enhances their reliability.



\section{Summary and Conclusions}\label{sec:conclusion}

We have categorized feature selection (FS) methods used by current DRSs into two groups: early FS and late FS. 
Being adaptive to the individual samples, late FS is better than early FS in terms of the recommendation performance. However, the former has a major limitation as its inability to practically reduce the activated embedding parameters.
We propose  \model, an \modelFullName method, which dynamically selects the most relevant features for each user-item interaction while significantly reducing the activated parameters in the embedding layer.
We achieved this through the integration of an \auxModel and \mainModel, utilizing collaborative training techniques to align their feature importance scoring. Our extensive experiments on benchmark datasets like Avazu and Criteo demonstrate that \model not only maintains the performance level of state-of-the-art adaptive feature selection methods but does so with parameter effiency, by approximately 37.5\%. These findings underscore the effectiveness of our approach in balancing efficiency and accuracy, making \model a feasible solution for large-scale recommender systems.





\bibliographystyle{IEEEtran}
\bibliography{sample-base}

\begin{IEEEbiography}[{\includegraphics[width=1in,height=1.25in,clip,keepaspectratio]{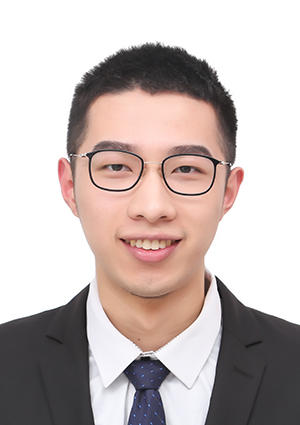}}]{Fan Hu}
received his PhD degree from Renmin University of China in 2025. His research interests include multi-modal retrieval, multi-modal LLM, and deep recommender systems.
\end{IEEEbiography}
\begin{IEEEbiography}[{\includegraphics[width=1in,height=1.25in,clip,keepaspectratio]{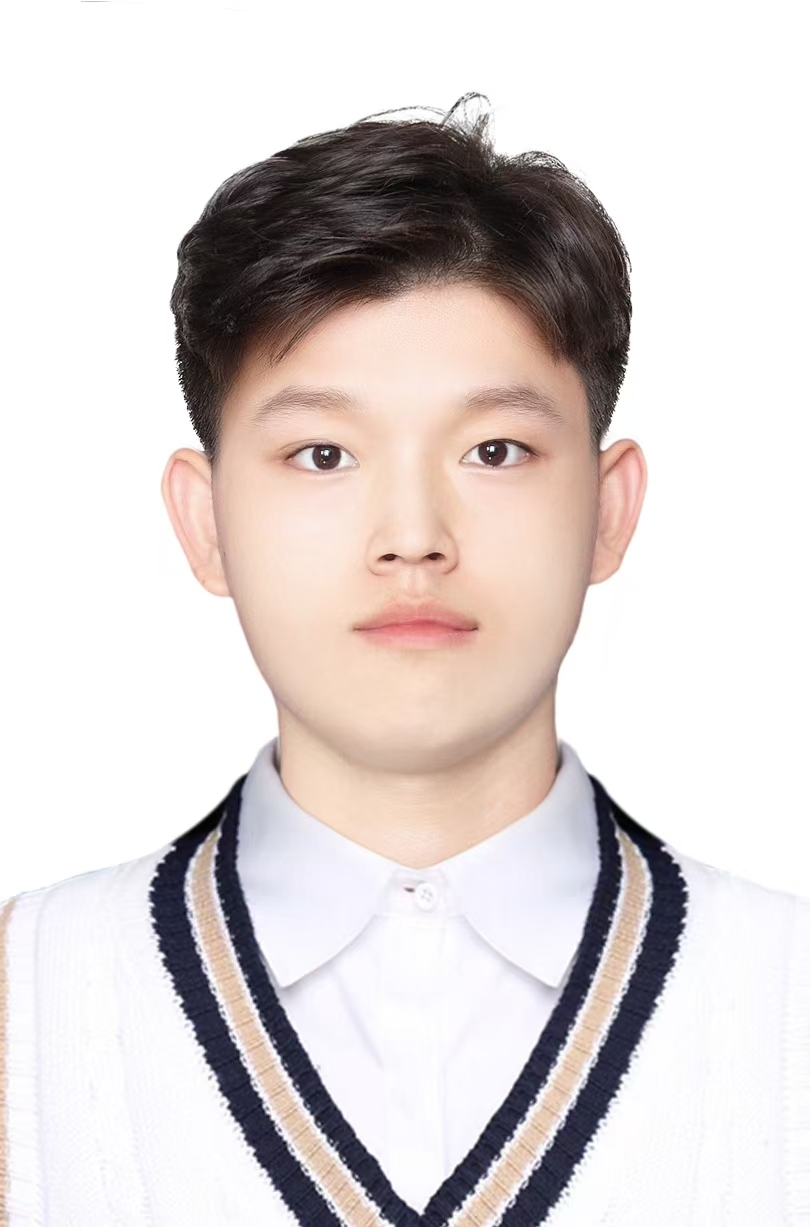}}]{Gaofeng Lu}
is currently pursuing his master’s degree in the University of Science and Technology of China. He is interning at Tencent. His research interests include LLM, VLM and recommender system.
\end{IEEEbiography}
\begin{IEEEbiography}[{\includegraphics[width=1in,height=1.25in,clip,keepaspectratio]{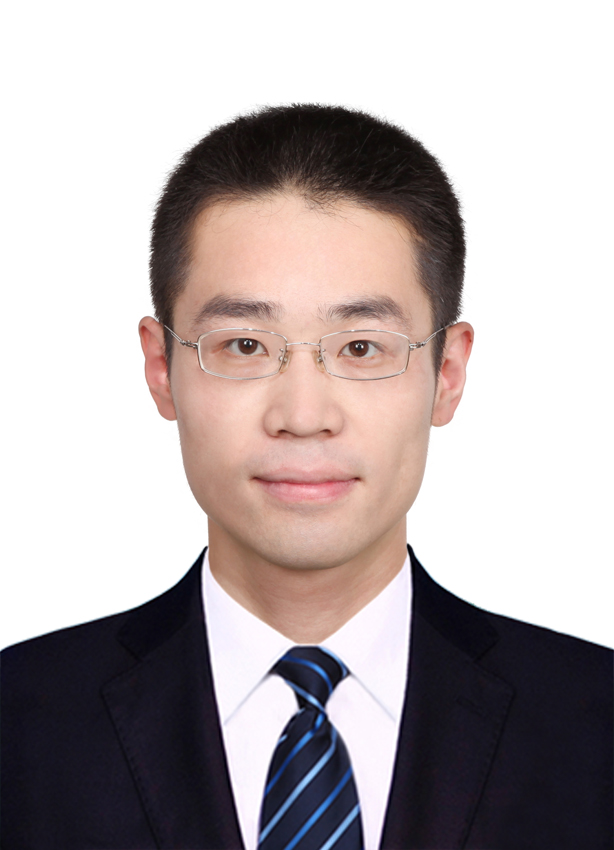}}]{Jun Chen}
obtained his PhD degree from École Centrale de Marseille, France. He is currently affiliated with Tencent Inc., where his research focuses on high-performance computing and the development of machine learning platforms.
\end{IEEEbiography}
\begin{IEEEbiography}[{\includegraphics[width=1in,height=1.25in,clip,keepaspectratio]{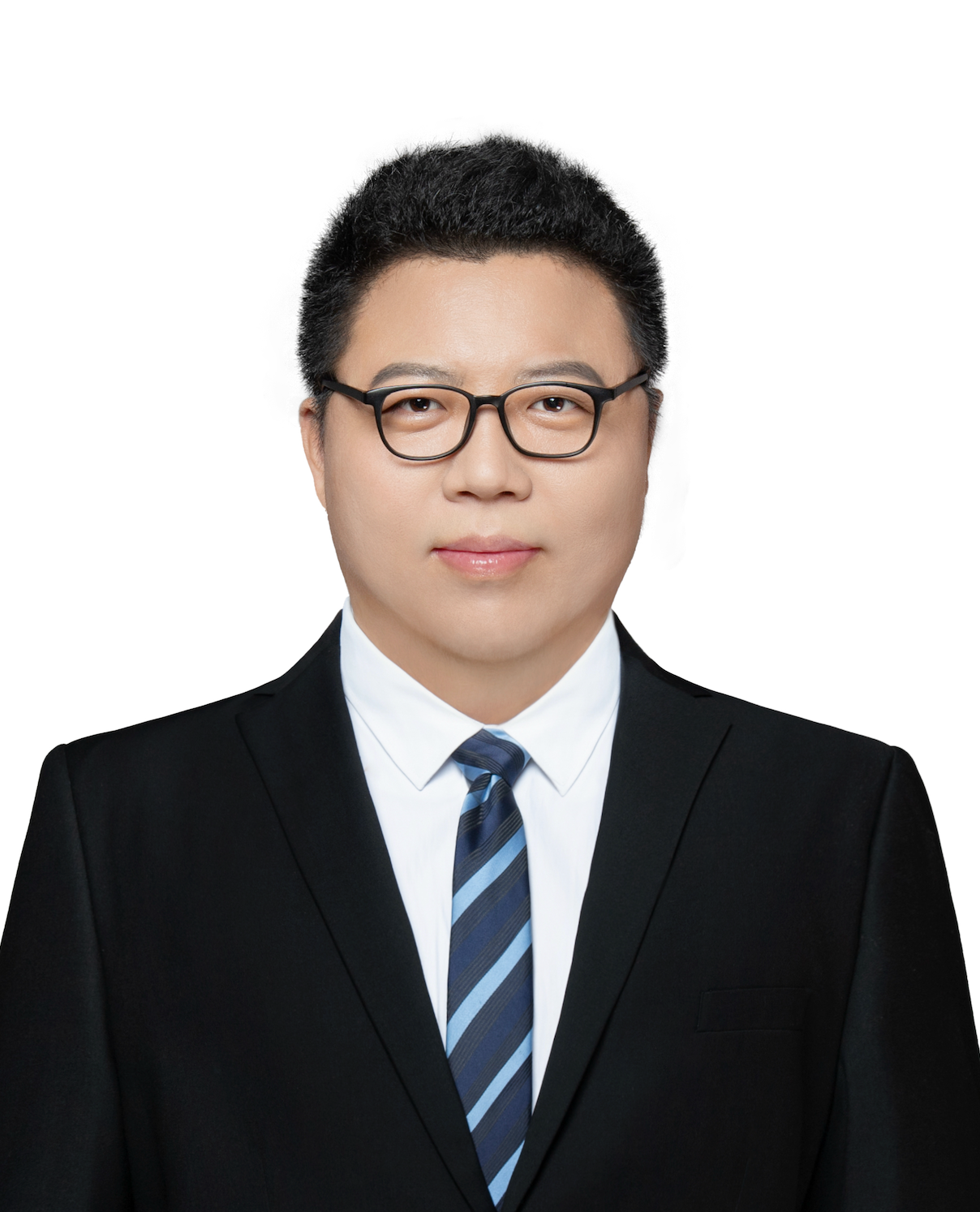}}]{Chaonan Guo}
  graduated from Harbin Institute of Technology and is currently engaged in development work related to recommendation systems, machine learning, and big data.
\end{IEEEbiography}
\begin{IEEEbiography}[{\includegraphics[width=1in,height=1.25in,clip,keepaspectratio]{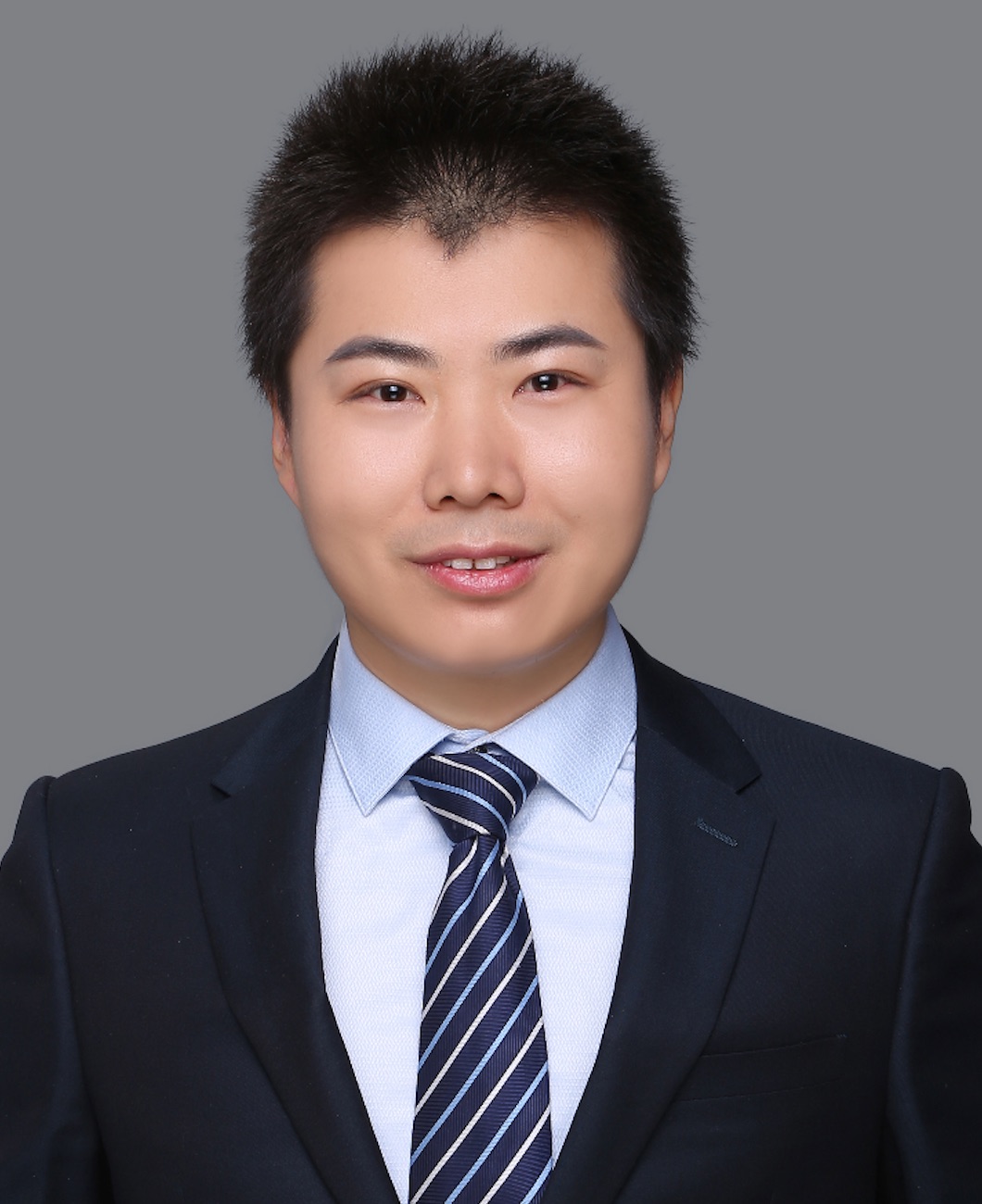}}]{Yuekui Yang}
 received his B.E. from Taiyuan University of Technology in 2006. He is currently pursuing an Eng.D. degree in Department of Computer Science and Technology, Tsinghua University. His research interests focus on recommendation systems and computational advertising. 
\end{IEEEbiography}
\begin{IEEEbiography}[{\includegraphics[width=1in,height=1.25in,clip,keepaspectratio]{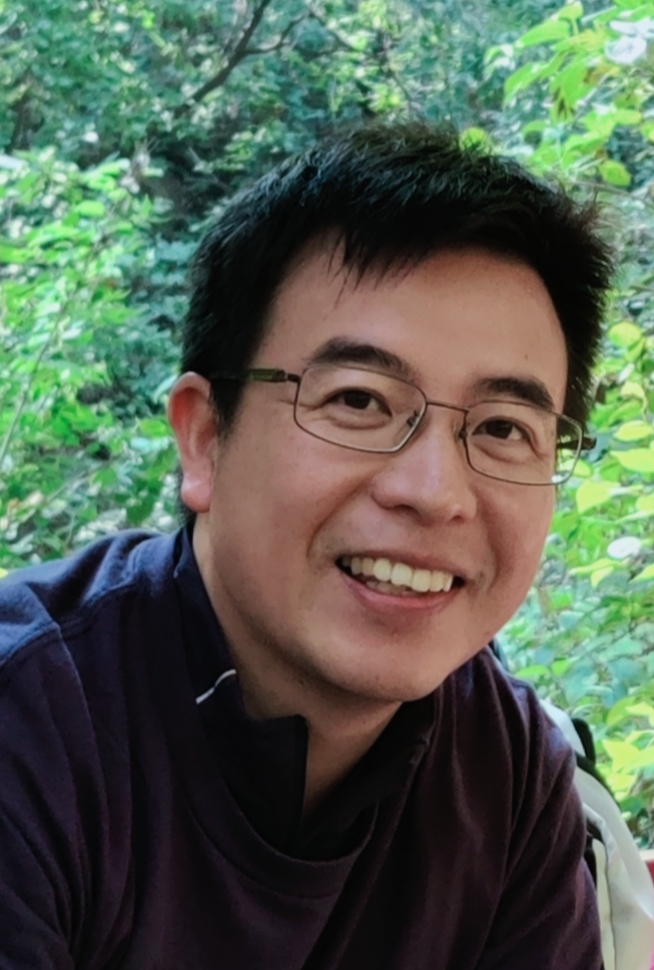}}]{Xirong Li}
(Member, IEEE) received the B.S. and M.E. degrees from Tsinghua University, Beijing, China, in 2005 and 2007, respectively, and the Ph.D. degree from the University of Amsterdam, Amsterdam, The Netherlands, in 2012, all in computer science. He is currently a Full Professor with the School of Information, Renmin University of China, Beijing. His research focuses on multimodal intelligence. He was recipient of CCF Science and Technology Award 2024, the ACMMM 2016 Grand Challenge Award, the ACM SIGMM Best Ph.D. Thesis Award 2013, the IEEE Transactions on Multimedia Prize Paper Award 2012, and the Best Paper Award of ACM CIVR 2010. He served as Program Co-Chair for Multimedia Modeling 2021 and Associate Editor for IET Computer Vision, and is serving as Associate Editor for ACM TOMM and the Multimedia Systems journal.
\end{IEEEbiography}



\vfill

\end{document}